\documentclass[onecollarge,runningheads,envcountsect]{svjour2}%{}

\smartqed 
\usepackage{setspace}
\usepackage{latexsym}
\usepackage{graphicx}
\usepackage{epsfig}
\usepackage{amssymb}
\usepackage{amsmath}
\usepackage{amsfonts}
\usepackage{dsfont}

\newtheorem{dfn}{Definition}[section]
\newtheorem{Rem}{Remark}[section]
\newtheorem{Th}{Theorem}[section]

\def\bN{{\bf N}}

\def\bd{{\bf d}}

\newfont {\car} {eufm10}
\newfont {\Lar} {eufb10}
\newfont {\lar} {eufm5}
\newfont {\ab} {msbm5}
\newfont {\AB} {msbm10}
\newfont {\Lab} {msbm10}

\def\beq{\begin{equation}}
\def\eeq{\end{equation}}

\def\bd0{{\bf 0}}

\journalname{Continuum Mechanics and Thermodynamics}

\begin{document}

\title{Thermodynamics of non-local materials: extra fluxes and internal powers }
\author{Mauro Fabrizio, Barbara Lazzari and Roberta Nibbi }
\institute{ \at Department  of  Mathematics University of Bologna \\
\email{fabrizio@dm.unibo.it,  lazzari@dm.unibo.it,  nibbi@dm.unibo.it } }
\authorrunning{M. Fabrizio, B.  Lazzari and R. Nibbi}
\titlerunning{Thermodynamics of non-local materials}
\date{}
\maketitle

\begin{abstract}
The most usual  formulation of the Laws of Thermodynamics turns out to be suitable for local or simple materials, while for non-local systems  there are  two different ways:
either modify this usual formulation by introducing suitable {extra fluxes}
or  express   the Laws of Thermodynamics in terms of internal powers directly, as we propose in this paper.
 
The first choice is subject to the criticism that the  vector fluxes  must be introduced a posteriori in order to obtain the compatibility with the Laws of Thermodynamics. On the contrary,   the formulation in terms of internal powers  is more general , because it is a priori defined on the basis of the constitutive equations. Besides
 it allows  to highlight, without ambiguity,
the contribution of the  internal powers in the variation of the thermodynamic potentials.
 
 Finally,  in this paper, we consider  some examples of non-local materials and  derive the  proper expressions of their internal powers from the power balance laws. 
\end{abstract}

\renewcommand{\theequation}{\thesection.\arabic{equation}} %
\setcounter{equation}{0}

\section{Introduction}
The modern approach to the thermodynamics of continuous systems, known as 
\emph{Rational Thermodynamics}, traces back to the works of Coleman, Gurtin,
Noll and Truesdell \cite{ColemanNoll1963}, \cite{Gurtin1981}, \cite{noll1958},
\cite{Truesdell1966}. It's our opinion that this important new approach in
the study of thermodynamic phenomena has achieved its formal stage in the
papers of Noll  \cite{Noll1972} and Coleman and Owen \cite{ColemanOwen1974}. As it is well known, this
innovative point of view distinguishes the field equations, which hold for
all systems, from the constitutive equations, which characterize the
materials. Moreover, according to this school, restrictions on the constitutive equations can be obtained from the Principles of
Thermodynamics.

In most of the works of the previously cited authors only \emph{simple
materials} are considered. These systems are defined by Truesdell (\cite%
{Truesdell1966}, pag.35) as those materials for which \emph{\
everything there is to know can be found out by performing experiments on
homogeneous motions of a body of those materials, from whatever state they
happen to find in it}. In the same book, Truesdell gives also a formal
definition of simple materials, in presence of purely mechanical and
isothermal processes, through the constitutive equation 
$
\mathbf{T}({x},t)=\hat{\mathbf{T}}({\mathbf{F}}^{t}({x},\cdot
);{x})
$
which links the stress tensor $\mathbf{T}$ to the history up to time $t$ of
the strain gradient ${\mathbf{F}}$, at any point ${x}$ of the body $%
\mathcal{B}$. 

A more rigorous and general definition of simple material can
be found in \cite{Noll1972} and \cite{ColemanOwen1974}.
These materials are always characterized by the fact that the First Law of
Thermodynamics is given by 
\begin{equation}  \label{simple1}
\rho \frac {d}{dt}e= \mathbf{T} \cdot \mathbf{L} - \nabla \cdot \mathbf{q} +
\rho r,
\end{equation}
where $\rho$ is the mass density, $e$ the internal energy, $\mathbf{L}$ the
velocity gradient, $\mathbf{q}$ the heat flux and $r$ the heat supply, while
the Second Law of Thermodynamics is formulated through the Clausius-Duhem
inequality 
\begin{equation}  \label{simple2}
\rho \frac {d}{dt}\eta \geq - \nabla \cdot \left( \frac{\mathbf{q}}{\theta} \right)+ \rho 
\frac{r}{\theta} ,
\end{equation}
$\eta$ being the specific entropy density and $\theta$ the absolute
temperature.

The expressions (\ref{simple1}) and (\ref{simple2}) of the First and Second
Law of Thermodynamics cannot be applied to those systems, called \emph{%
non-local} or \emph{non-simple materials}, for which the constitutive
equations for $\mathbf{T}$ and $\mathbf{q}$ depend on the gradients of
order higher than the first of the thermo-kinetic variables $( \mathbf{F},
\theta)$.

In order to understand how it is possible to give a formulation of the First
Law of Thermodynamics for a wider class of materials, we consider the heat
balance equation \cite{ColemanOwen1974} 
\begin{equation}
\rho h=-\nabla \cdot \mathbf{q}+\rho r,  \label{simple4}
\end{equation}%
where $h$ is the rate at which the heat is absorbed per unit mass. By
comparing (\ref{simple4}) with (\ref{simple1}), the First Law of
Thermodynamics for simple materials can be rewritten as follows 
\begin{equation}
\rho \frac{d}{dt}e=\mathbf{T}\cdot \mathbf{L}+\rho h.  \label{simple3}
\end{equation}%
Moreover, following many classical texts of Thermodynamics (Truesdell \cite%
{Truesdell1991}, Gurtin \cite{Gurtin1972}, Germain \cite{Germain1973,Germain1973art} , etc.), we introduce the \textit{internal mechanical power} ${\mathcal{P}}_{m}^{i}$ and the \emph{internal heat power} $%
{\mathcal{P}}_{h}^{i}=\rho h$. Then the balance equation for the energy (\ref%
{simple3}) assumes the classical form 
\begin{equation}
\rho \frac{d}{dt}e={\mathcal{P}}_{m}^{i}+{\mathcal{P}}_{h}^{i},
\label{simple5}
\end{equation}%
which is more general and can also be applied to non-simple materials.
Obviously it coincides with (\ref{simple3}) when 
\begin{equation}\label{Ns1}
{\mathcal{P}}_{m}^{i}=\mathbf{T}\cdot \mathbf{L},
\end{equation}%
but this choice, which is surely correct for simple materials, cannot be
assumed in general. For this reason Dunn and Serrin \cite{DunnSerrin1985}
suggested to change the formulation (\ref{simple1}) of the First Law of
Thermodynamics with the new expression%
\begin{footnote}{Only recently it has
been observed that expression (\ref{simple6}) and the introduction of the
extra-flux $\bN$ in the First Law of Thermodynamics had been already
proposed in 1982 by T. Manacorda \cite{Manacorda1982}, see also
\cite{Manacorda1986}.}\end{footnote}
\begin{equation}  \label{simple6}
\rho \frac {d}{dt}e= \mathbf{T} \cdot \mathbf{L} - \nabla \cdot \mathbf{N} -
\nabla \cdot \mathbf{q} + \rho r,
\end{equation}
where $\mathbf{N}$ is called \textit{interstitial work flux}. Expression (%
\ref{simple6}) has been actually introduced by the authors for some non-simple materials 
including 
the elastic materials of grade three and 
the subclass of materials of Korteweg type, but its framework is more general (\cite{FriedGurtin2006,Mehrabadi2005}).
% \cite{Mehrabadi2005}.

In this paper we start from the statement that the representations (\ref%
{simple1}) or (\ref{simple3}) of the First Law of Thermodynamics cannot have
universal character and propose the more general classical formulation (\ref%
{simple5}), where the internal mechanical power ${\mathcal{P}}_{m}^{i}$ will
assume different expressions depending on the materials under study.

It is also interesting to observe that a comparison between Dunn and
Serrin's formulation (\ref{simple6}) and the classical one (\ref{simple5})
gives 
\begin{equation}
{\mathcal{P}}_{m}^{i}=\mathbf{T}\cdot \mathbf{L}-\nabla \cdot \mathbf{N},
\label{simple7}
\end{equation}%
which turns out to be an indirect way to define ${\mathcal{P}}_{m}^{i}$. Let
us consider, as an example, the constitutive equation for the stress tensor
of a second grade material, i.e. 
\begin{equation*}
\mathbf{T}=\mathbf{T}_{2}+\nabla \cdot \mathbf{T}_{3}, % \label{simple8}
\end{equation*}%
where $\mathbf{T}_{2}$ and $\mathbf{T}_{3}$ are respectively second and
third order tensors. The internal mechanical power is given by (\cite%
{CardonaForest1999,FriedGurtin2006,Germain1973art}) 
\begin{equation}
{\mathcal{P}}_{m}^{i}=\mathbf{T}_{2}\cdot \nabla \mathbf{v}-\mathbf{T}%
_{3}\cdot \nabla (\nabla  \mathbf{v}). \label{simple9}
\end{equation}%
Since 
\begin{equation*}
\mathbf{T}_{3}\cdot  \nabla (\nabla  \mathbf{v})=\nabla \cdot \left( \mathbf{T}%
_{3}\nabla \mathbf{v}\right) -(\nabla \cdot \mathbf{T}_{3})\cdot \nabla 
\mathbf{v},
\end{equation*}%
expression (\ref{simple7}) turns out to be equivalent to (\ref{simple9}) introducing the vector\begin{equation*}
\mathbf{N}=\mathbf{T}_{3}\nabla \mathbf{v}. % \label{simple10}
\end{equation*}%
Nevertheless, we prefer the representation of the internal mechanical power (%
\ref{simple9}), since it is in agreement with the expression of the internal
power as a differential form and since (\ref{simple7}) results a
\textquotedblright hybrid\textquotedblright\ form because it contains a term
of ambiguous differential type, $\mathbf{T}\cdot \mathbf{L}$, and a flux
term.

Similar arguments can be applied to many non-local mechanical materials, to
the study of the phase transitions of superconductors and superfluids, to
the transition of a binary mixture satisfying the Cahn-Hilliard equations
and, in electromagnetism, to dielectric media with quadrupoles.

In literature some authors maintain the expression (\ref{simple1}) for the
First Law of Thermodynamics and, in presence of the above non-simple
materials, modify the Clausius-Duhem inequality (\ref{simple2}). They
consider, as suggested in 1967 by M\"{u}ller \cite{Muller1967}, the
following inequality 
\begin{equation}
\rho \frac{d}{dt}\eta \geq -\nabla \cdot \mathbf{\Phi }+\rho \frac{r}{\theta 
}\,,  \label{simple12}
\end{equation}%
where the vector $\mathbf{\Phi}$, still called entropy flux, can be defined
by 
\begin{equation}
\mathbf{\Phi} = \mathbf{\Phi} ^{\prime }+\frac{\mathbf{q}}{\theta },
\label{simple13}
\end{equation}%
$ \mathbf{\Phi} ^{\prime}$ being the \textit{entropy extra-flux}. 

This
approach is useful only  in presence of non-local effects on the heat flux.
Instead it is not fruitful in the study of materials with constitutive equations describing non-local behaviors in mechanics, electromagnetism, chemistry, etc..
For these materials the Second Law assumes the classical form (\ref{simple2}), while we need to provide a suitable expression for the (mechanical, chemical, electromagnetic, etc.) internal powers to use in the First Law.

On the other hand, some thermal effects are well described by non-local
mo\-dels for which the Clausius-Duhem inequality (\ref{simple2}) must be
substituted by (\ref{simple12}). An example is the Guyer-Krumhansl equation 
\cite{Guyer-Krumhansl1966a,Guyer-Krumhansl1966b},
 \begin{equation}
\frac{d}{dt}\mathbf{q}+\frac{\mathbf{q}}{\tau }=-\kappa (\theta )\nabla
\theta +\alpha (\nabla \cdot \nabla \mathbf{q}+2\nabla \nabla \cdot \mathbf{q%
}),  \label{GK}
\end{equation}%
 which has beeen derived by solving the
linearized Boltzmann equation of phonon gas hydrodynamics.
Here $\tau $, $\kappa $ and $\alpha $ are suitable constitutive
coefficients and this model,  as proved in \cite{Cimmelli2007},  satisfies the inequality (\ref{simple12})
  by choosing 
\begin{equation*}
 \mathbf{\Phi} ^{\prime}=\frac{\alpha }{\kappa }((\nabla \mathbf{q})\mathbf{q%
}+2\mathbf{q}\nabla \cdot \mathbf{q}).
\end{equation*}%

The presence of the extra-fluxes $\mathbf{N}$ in (\ref{simple7}) and $%
 \mathbf{\Phi}^{\prime}$ in (\ref{simple12})-(\ref{simple13}) opens to the
criticism of a formulation of the Principles of Thermodynamics which turns
out to be not universal, since these vectors must be defined a posteriori.

As we have already observed, a general formulation for the First Law of
Thermodynamics is given by (\ref{simple5}), where the internal power
densities ${\mathcal{P}}^i_m $ and ${\mathcal{P}}^i_{h}$ are a priori
defined on the basis of the constitutive equations.
In order to get a similar general expression for the Second Law of
Thermodynamics, we 
introduce the entropy balance equation 
\begin{equation}
\rho \frac{h}{\theta }=-\frac{1}{\theta }\nabla \cdot {\mathbf{q}}+\rho 
\frac{r}{\theta },  \label{simple15}
\end{equation}%
which can be derived multiplying equation (\ref{simple4}) by the coldness $\theta^{-1}$. 

This equation distinguishes between the contributions of
the internal and external actions because (\ref{simple15}) can be
rewritten as 
\begin{equation}
\rho \frac{h}{\theta }+\frac{1}{\theta ^{2}}\mathbf{q}\cdot \nabla \theta
=-\nabla \cdot \left(\frac{\mathbf{q}}{\theta }\right)+\rho \frac{r}{\theta }.
\label{simple16}
\end{equation}%
For simple materials in the heat flux (i.e. for  materials with  a local constitutive equation for the heat flux), we define  \textit{internal entropy action}  the left-hand side of (\ref{simple16}),
which is  a  differential form, and  \textit{external entropy action}  the right-hand side of  (\ref{simple16}), i.e.
\begin{equation}
{\mathcal{A}}_{en}^{i}=\rho \frac{h}{\theta }+\frac{1}{\theta ^{2}}\mathbf{q}%
\cdot \nabla \theta \,, \qquad   {\mathcal{A}}_{en}^{e}=-\nabla \cdot \frac{\mathbf{q}}{\theta }+\rho \frac{r}{\theta }.  \label{simple17}
\end{equation}%
Therefore, thanks to  (\ref{simple16}),   the usual Clausius-Duhem inequality (\ref{simple2}) assumes the equivalent form 
\begin{equation}
\rho \frac{d}{dt}\eta \geq 
\rho \frac{h}{\theta }+\frac{1}{\theta ^{2}}\mathbf{q}\cdot \nabla \theta 
 \label{simple19}
\end{equation}%
or, by virtue of  (\ref{simple17}), it can be expressed only by means of
the internal entropy action as follows
\begin{equation}
\rho \frac{d}{dt}\eta \geq 
{\mathcal{A}}_{en}^{i} . \label{simple191}
\end{equation}%
We observe that the expression (\ref{simple19}) holds
only for simple materials in the heat flux, while the relation  (\ref{simple191})  is consistent also  with the modified Clausius Duhem inequality  (\ref{simple12})  when  the internal entropy action is given by 
\begin{equation}
{\mathcal{A}}_{en}^{i}=\rho \frac{h}{\theta }+\frac{1}{\theta ^{2}}\mathbf{q}%
\cdot \nabla \theta -\nabla \cdot  \mathbf{ \Phi}^{\prime }.  \label{simple171}
\end{equation}%
It is therefore clear that it is necessary to give a proper definition of internal and external entropy action.
In fact, if for example we consider a heat
conductor of Guyer-Krumhansl type  (\ref{GK}) with $\displaystyle\kappa
(\theta )=\frac{c}{\theta ^{2}}$, the entropy balance equation
%$$\rho \frac{h}{\theta }=-\nabla \cdot\left[ \frac{\mathbf{q}}{\theta }\right] -\frac{1}{\theta ^{2}}\mathbf{q}\cdot \nabla \theta+\rho \frac{r}{\theta }$$
 becomes 
 $$\rho \frac{h}{\theta }=-\nabla \cdot\left[ \frac{\mathbf{q}}{\theta }\right] + \frac 1c\mathbf{q}\cdot \left[ \frac{d}{dt}\mathbf{q}+\frac{\mathbf{q}}{\tau }-\alpha (\nabla \cdot \nabla \mathbf{q}+2\nabla \nabla \cdot \mathbf{q})\right]+\rho \frac{r}{\theta }\,.
$$
Since
$$
\mathbf{q}\cdot \left( \nabla \cdot \nabla \mathbf{q}+2\nabla \nabla \cdot \mathbf{q}\right)=
\nabla \cdot\left[(\nabla \mathbf{q})\mathbf{q}+2\mathbf{q}\nabla \cdot \mathbf{q}\right]
-|\nabla \mathbf{q}|^{2}-2|\nabla \cdot \mathbf{q}|^{2}\,, 
$$
we obtain 
\begin{equation*}
\rho \frac{h}{\theta }-\frac{d}{dt}\left( \frac{\mathbf{q}^{2}}{2c}\right)
-\frac{1}{c\tau }\mathbf{q}^{2}-\frac{\alpha }{c}\left( |\nabla \mathbf{q}%
|^{2}+2|\nabla \cdot \mathbf{q}|^{2}\right)  
=-\nabla \cdot \left[ \frac{\mathbf{q}}{\theta }+\frac{\alpha }{c}\left(
(\nabla \mathbf{q})\mathbf{q}+2\mathbf{q}\nabla \cdot \mathbf{q}\right) %
\right] +\rho \frac{r}{\theta }.
\end{equation*}%
It follows that, for such material, the internal entropy action is defined by
\begin{equation*}
{\mathcal{A}}_{en}^{i}=\rho \frac{h}{\theta }-\frac{d}{dt}\left( \frac{%
\mathbf{q}^{2}}{2c}\right) -\frac{1}{c\tau }\mathbf{q}^{2}-\frac{\alpha }{c}%
\left( |\nabla \mathbf{q}|^{2}+2|\nabla \cdot \mathbf{q}|^{2}\right) ,
%\label{simple20}
\end{equation*}%
while the external entropy action is given by
\begin{equation}
{\mathcal{A}}_{en}^{e}=\rho \frac{r}{\theta }-\nabla \cdot \left[ \frac{%
\mathbf{q}}{\theta }+\frac{\alpha }{c}\left( (\nabla \mathbf{q})\mathbf{q}+2%
\mathbf{q}\nabla \cdot \mathbf{q}\right) \right] .  \label{simple21}
\end{equation}%
Finally, in this paper, we suggest to describe the evolution of a
thermodynamic system through the  virtual mechanical power and entropy action. To
this end, it is important to introduce the domain $\Omega \subset 
\hbox{\AB
{R}}^{3}$ on which are defined the fields and the processes of the virtual
velocity $\widetilde{\mathbf{v}}$ and of the virtual temperature $\widetilde{%
\theta }$. By denoting the internal and external  virtual mechanical powers
respectively with $\widetilde{{\mathcal{P}}}_{m}^{i},\widetilde{{\mathcal{P}}%
}_{m}^{e}$ and the internal and external virtual entropy  actions with $%
\widetilde{{\mathcal{A}}}_{en}^{i},\widetilde{{\mathcal{A}}}_{en}^{e}$, we
are now able to write the balance equation for the virtual  mechanical power 
\begin{equation*}
\int_{\Omega }\rho \frac{d}{dt}\mathbf{v}\cdot \widetilde{\mathbf{v}}{\,d}%
x+\int_{\Omega }\widetilde{{\mathcal{P}}}_{m}^{i}{\,d}x=\int_{\Omega }%
\widetilde{{\mathcal{P}}}_{m}^{e}{\,d}x  %\label{simple22}
\end{equation*}%
and the balance equation for the virtual entropy  action
\begin{equation*}
\int_{\Omega }\widetilde{{\mathcal{A}}}_{en}^{i}{\,d}x=\int_{\Omega }%
\widetilde{{\mathcal{A}}}_{en}^{e}{\,d}x.  %\label{simple23}
\end{equation*}%
In presence of simple materials, the previous equations  become 
\begin{eqnarray*}
\displaystyle{\int_{\Omega }\rho \frac{d}{dt}\mathbf{v}\cdot \widetilde{\mathbf{v}}{\,d}%
x+\int_{\Omega }\mathbf{T}\cdot \nabla \widetilde{\mathbf{v}}{\,d}%
x}&=&\displaystyle{\int_{\Omega }\left[ \nabla \cdot (\mathbf{T}\widetilde{\mathbf{v}})+\rho 
\mathbf{f}\cdot \widetilde{\mathbf{v}}\right] {\,d}x,}
\\
\displaystyle{\int_{\Omega }\left[ \rho \frac{h}{\widetilde{\theta }}-\frac{1}{\widetilde{%
\theta }^{2}}\mathbf{q}\cdot \nabla \widetilde{\theta }\right] {\,d}%
x}&=&\displaystyle{\int_{\Omega }\left[ -\nabla \cdot \left( \frac{\mathbf{q}}{\widetilde{%
\theta }}\right) +\rho \frac{r}{\widetilde{\theta }}\right] {\,d}x,}
\end{eqnarray*}
where $\mathbf{f}$\  denotes the external forces and the symmetry the the stress tensor $\mathbf{T}$ is assumed.

%%%%%%%%%%%%%%%%%
%SISTEMA DINAMICO
%%%%%%%%%%%%%%%%

\setcounter{equation}{0}
\section{Definition of thermo-mechanical material as  dynamic system}
The thermo-mechanical properties of a material are based on the notion of 
\textit{state} and \textit{process}. Here we provide these notions
following Noll \cite{Noll1972} and Coleman and Owen \cite{ColemanOwen1974} ( see also
\cite{FMlibro}).

We consider a body occupying the placement $\mathcal{B}$.
For any material point $X\in \mathcal{B}$, a \textit{ thermo-mechanical process} 
$P$ in $X$%
%\begin{footnote}{From now on, whenever no ambiguity arises, the dependence on $X$ is understood and not written. }\end{footnote}%
\negthinspace \negthinspace \negthinspace , of du\-ra\-tion $d_{P}>0$, is a
piecewise continuous function on $[0,d_{P})$ with values in $E$, open and
connected subset of a vectorial space.

The notation $P_{[t_1,t_2)}$ denotes the restriction of $P$ to $%
[t_1,t_2)\subset [0,d_p)$. In particular, we denote by $P_t$ the restriction
of $P$ to the interval $[0,t)$, $t\leq d_P$. Let $P_1$ and $P_2$ be two
processes of duration $d_{P_1}$ and $d_{P_2}$, the \textit{composition} $%
P_1\star P_2$ of $P_2$ with $P_1$ is defined as 
\begin{equation*}
P_1\star P_2=\left\{ 
\begin{array}{ll}
P_1(t), & t\in[0,d_{P_1}) \\ 
P_2(t- d_{P_1}), & t\in[d_{P_1},d_{P_1}+d_{P_2})%
\end{array}
\right.
\end{equation*}
We can therefore give the following definition of  thermo-mechanical material
as dynamic system.
\begin{dfn}
\label{def1} \rm{{\ Let $\mathcal{B}_t$ the placement occupied by the
body at time $t$. } 
{A  thermo-mechanical system, at any $t$ and $x\in \mathcal{B}_{t}$,
is a set $\{\Pi ,\widetilde{\mathcal{V}}_{t},\Sigma ,\widehat{\chi },{U}\}$
such that }
\begin{enumerate}
\item $\Pi$ is the space of the  thermo-mechanical processes $P$
satisfying the following properties: 
\newcounter{storia} 
\begin{list}
{\roman{storia} ) }{\usecounter{storia}}
    \item if $P\in\Pi$, then $P_{[t_1,t_2)}\in \Pi$ for every $[t_1,t_2)\subset
    [0,d_p)$,
    \item if $P_1$, $P_2\in \Pi$, then $P_1\star P_2 \in \Pi$,
  \end{list}
\item $\widetilde{\mathcal{V}}_t$ is the space of the virtual
 thermo-mechanical processes $\widetilde{\nu}_t$, i.e. the space of the
processes that may be specified independently of the actual evolution of the
system at some arbitrarily chosen but fixed time $t$,
\item $\Sigma$ is an abstract set, called \textit{state space},
whose elements $\sigma$ are called \textit{states}, 
\item $\widehat{\chi}:\Sigma\times\Pi\to\Sigma$ is a map, called 
\textit{state transition function} which  assigns,  to each state $\sigma$ and process 
$P$, the state at the end of the process and satisfies the following
relation 
\begin{equation*}
\widehat{\chi}(\sigma, P_1\star P_2)= \widehat{\chi}(\widehat{\chi}(\sigma,
P_1),P_2)\;,\quad \sigma\in \Sigma\;, P_1,P_2\in \Pi,
\end{equation*}
\item for any state $\sigma$,  process $P$ of duration $d_P$, time 
$t \in [0, d_P)$ and virtual process $\widetilde{\nu}_t\in \widetilde{%
\mathcal{V}}_t$, it is defined the \textit{response function} at time $t$ 
\begin{equation*}
\widehat{U}_t:\Sigma \times E \times\widetilde{\mathcal{V}}_t\to\mathbb{R }%
\times \mathbb{R }
\end{equation*}
which assigns the internal virtual mechanical  power ${\widetilde{\mathcal{P}}%
}^i_m$ and the internal virtual entropy  action$\widetilde{\mathcal{A}}%
^i_{en}$, namely 
\begin{equation*}
{U}(t)=\widehat{U}_t(\sigma_t, P(t), \widetilde{\nu}_t) = ({\widetilde{%
\mathcal{P}}}^i_m, {\widetilde{\mathcal{A}}}^i_{en})_t\quad t \in [0, d_P),
\end{equation*}
where $\sigma_t= \widehat{\chi}(\sigma, P_t)$. 
\end{enumerate}}
\end{dfn}
\begin{Rem}
\rm{{Because of the linear dependence of the virtual power and action on the
virtual processes, the function ${U}(t)$ is linear with respect to $%
\widetilde{\nu}_t$. } }
\end{Rem}
\begin{Rem}
\rm{{\ The knowledge of ${\widetilde{%
\mathcal{P}}}^i_m$ and $ {\widetilde{\mathcal{A}}}^i_{en}$ allows to define the internal
mechanical power ${{\mathcal{P}}}^i_m$ and the internal entropy action ${%
\mathcal{A}}^i_{en}$. Moreover, if we restrict $\widetilde{%
\mathcal{A}}^i_{en}$ to spatial homogeneous thermal processes, we can also
define $h$
and the heat
flux power $\displaystyle{\mathcal{P}}_q={\mathcal{A}}^i_{en}- \rho \frac{ h%
}{\theta} $.} }
\end{Rem}

%%%%%%%%%%%%%%%%%
%PRINCIPI TERMODINAMICA
%%%%%%%%%%%%%%%%
\setcounter{equation}{0}

\section{Thermodynamic Laws}
In order to introduce the Thermodynamic Laws, following \cite%
{ColemanOwen1974}, we introduce the notion of cycle.
\begin{dfn}
\label{ciclo}\rm{{\ A pair $(\sigma,P)\in\Sigma\times\Pi$ is called a
cycle if\, $\widehat{\chi}(\sigma, P)=\sigma$. } }
\end{dfn}
 \textbf{First Law of Thermodynamics.} \textit{In any cycle $%
(\sigma,P)$ the sum of the rate at which the heat is absorbed by the body and
the power of the internal forces vanishes, i.e. 
\begin{equation*} % \label{primalegge}
\oint_0^{d_P}\left\{ \rho h(\sigma_t,P(t))+ \mathcal{P}_m^i(\sigma_t,P(t))%
\right\}\,d\,t=0.
\end{equation*}
}
A
 consequence of the First Law of
Thermodynamics is the following theorem
\begin{Th}
The First Law of Thermodynamics implies the existence of a unique (up to an
additive constant) continuous function $e:\Sigma\to\mathbb{R}$, called
internal energy, such that, for any pair of states $\sigma_1$, $\sigma_2$
and process $P$ with $\widehat{\chi}(\sigma_1, P)=\sigma_2$, we have 
\begin{equation*}  %\label{energiainterna1}
\rho e(\sigma_2)- \rho e(\sigma_1)=\int_0^{d_P}\left\{ \rho
h(\sigma_t,P(t))+ \mathcal{P}_m^i(\sigma_t,P(t))\right\}\,d\,t.
\end{equation*}
Moreover, under suitable hypotheses of regularity on the constitutive
functionals, 
\begin{equation}  \label{energiainterna2}
\rho \frac {d}{dt}e(\sigma_t)= \rho h(\sigma_t,P(t))+ \mathcal{P}%
_m^i(\sigma_t,P(t))
\end{equation}
for any instant $t$ of continuity of the right-hand term in \rm{(\ref{energiainterna2})}.
\end{Th}
\noindent \textbf{Second Law of Thermodynamics.} \textit{In any cycle $%
(\sigma,P)$ the internal entropy action is non-positive, i.e.
\begin{equation*}  %\label{ secondalegge}
\oint_0^{d_P} \mathcal{A}_{en}^i(\sigma_t,P(t))\,d\,t\leq0.
\end{equation*}
}
As a consequence of the Second Law of Thermodynamics we get % (see \cite{ColemanOwen1974})
the following theorem
\begin{Th}
The Second Law of Thermodynamics implies the existence of at least one upper
semicontinuous function $\eta:\Sigma\to\mathbb{R}$, called entropy, such
that, for any pair of states $\sigma_1$, $\sigma_2$ and process $P$ with $%
\widehat{\chi}(\sigma_1, P)=\sigma_2$, we have 
\begin{equation*}  %\label{ entropia 1}
\rho \eta(\sigma_2)- \rho \eta(\sigma_1)\geq\int_0^{d_P} \mathcal{A}%
_{en}^i(\sigma_t,P(t))\,d\,t.
\end{equation*}
Moreover, under suitable hypotheses of regularity on the constitutive
functionals, 
\begin{equation}  \label{entropia2}
\rho \frac {d}{dt}\eta(\sigma_t)\geq\mathcal{A}_{en}^i(\sigma_t,P(t))
\end{equation}
for any instant $t$ of continuity of the right-hand term in \rm{(\ref{entropia2})}.
\end{Th}
When only isothermal processes are involved, the two Laws of Thermodynamics
are replaced by the Dissipation Principle of mechanical work.

\noindent \textbf{Dissipation Principle.} \textit{In any cycle $(\sigma ,P)$ the internal mechanical power is non-negative, i.e.
\begin{equation*}
\oint_{0}^{d_{P}}\mathcal{P}_{m}^{i}(\sigma _{t},P(t))\,d\,t\geq 0.
%\label{secondalegge1}
\end{equation*}%
}
A consequence of the Dissipation Principle %(see \cite{ColemanOwen1974})
is the following
theorem
\begin{Th}
The Dissipation Principle implies the existence of at least one lower
semicontinuous function $\psi:\Sigma\to\mathbb{R}$, called free energy, such
that, for any pair of states $\sigma_1$, $\sigma_2$ and process $P$ with $%
\widehat{\chi}(\sigma_1, P)=\sigma_2$,  we have 
\begin{equation*}  %\label{ energia libera 1}
\rho \psi(\sigma_2)- \rho \psi(\sigma_1)\leq\int_0^{d_P} \mathcal{P}%
_m^i(\sigma_t,P(t))\,d\,t.
\end{equation*}
Moreover, under suitable hypotheses of regularity on the constitutive
functionals, 
\begin{equation}  \label{ energia libera 2}
\rho \frac {d}{dt}\psi(\sigma_t)\leq \mathcal{P}_m^i(\sigma_t,P(t))
\end{equation}
for any instant $t$ of continuity of the right-hand term in \rm{(\ref{ energia libera 2})}.
\end{Th}

%%%%%%%%%%%%%%%%%
%POTENZA VIRTUALE
%%%%%%%%%%%%%%%%
\setcounter{equation}{0}

\section{Virtual powers and  thermo-mechanical materials}

For a  thermo-mechanical 
system, we define the field of the generalized
virtual motions at time $t$ as the pair 
\begin{equation*}
\widetilde{\nu }_{t}(\cdot )=\left( \widetilde{\mathbf{v}}_t(\cdot ),%
\frac{1}{\widetilde{\theta }_t(\cdot )}\right) ,
\end{equation*}%
where $\widetilde{\mathbf{v}}_t(\cdot )$ and  $\displaystyle{\frac{1}{\widetilde{\theta }_t(\cdot )} } $ denote
respectively the vector field of the
virtual velocities and the
 virtual coldness on $%
\mathcal{B}_{t}$.

The virtual powers of the internal forces at any $x$ in $%
\mathcal{B}_{t}$ are taken as  linear forms in the gradients of the virtual
motions $\widetilde{\nu }_{t}$ evaluated at $x$. For a large class of
materials we can represent the internal  virtual  mechanical power as 
\begin{equation}
\widetilde{\mathcal{P}}_{m}^{i}(t,x)=\sum_{i\geq 1}{{\mathbf{T}}_{i+1}(t,x)\cdot
\nabla ^{(i)}\widetilde{\mathbf{v}}_t(x)}  \label{pim}
\end{equation}%
where $\nabla ^{(n)}$ denotes the gradient of order $n$ and $\mathbf{T}_{k}$
is a tensor of order $k$. Moreover, the internal  virtual entropy action is
defined by 
\begin{equation}
\widetilde{\mathcal{A}}_{en}^{i}(t,x)=\frac{1}{\widetilde{\theta }_t(x)}%
\varrho (t,x) h(t,x)-\sum_{i\geq 1}{{\mathbf{q}}_{i}(t,x)\cdot \nabla ^{(i)}\left[ 
\frac{1}{\widetilde{\theta }_t(x)}\right] }  \label{pie}
\end{equation}%
where $\mathbf{q}_{i}$ is a tensor of order $i$. Finally,  the virtual mechanical power  and entropy action of the external and contact forces are
respectively given by 
\begin{eqnarray*}
\displaystyle{\widetilde{\mathcal{P}}_{m}^{e}(t,x)}&=&\displaystyle{\nabla \cdot \left[ \sum_{i\geq 1}{{%
\mathbf{T}}_{i+1}(t,x)\nabla ^{(i-1)}\widetilde{\mathbf{v}}_t(x)}\right] +{\rho (t,x)
\mathbf{f}(t,x)}\cdot \widetilde{\mathbf{v}}_t(x),}%  \label{pem}
\\
\displaystyle{\widetilde{\mathcal{A}}_{en}^{e}(t,x)}&=&\displaystyle{-\nabla \cdot \left( \sum_{i\geq 1}{{%
\mathbf{q}}_{i}(t,x)\nabla ^{(i-1)}\left[ \frac{1}{\widetilde{\theta }_t(x)}%
\right] }\right) +\rho(t,x) r(t,x)\frac{1}{\widetilde{\theta }_t(x)}.}%  \label{pee}
\end{eqnarray*}
The principle of virtual power is the requirement that the
external and internal virtual powers and entropy actions be balanced, i.e. 
\begin{equation}
\varrho (t,x)\frac{d}{dt}{\mathbf{v}}(t,x)\cdot \widetilde{\mathbf{v}}_t(x)+%
\widetilde{\mathcal{P}}_{m}^{i}(t,x)=\widetilde{\mathcal{P}}_{m}^{e}(t,x)\,,%
\qquad \widetilde{\mathcal{A}}_{en}^{i}(t,x)=\widetilde{\mathcal{A}}%
_{en}^{e}(t,x)  \label{PVP}
\end{equation}%
for any generalized virtual motion.

From now on, whenever no ambiguity arises, the dependence on $x$ and $t$
will be understood and not written.
\begin{dfn}
\label{semplice}\rm{{A  thermo-mechanical material is called a \emph{simple
material } if 
\begin{equation*}
{\mathbf{T}}_{i+1}={\mathbf{0}}\,,\quad {\mathbf{q}}_{i}={\mathbf{0}}%
\,,\qquad i\geq 2
\end{equation*}%
for any time $t$ and $x\in \mathcal{B}_{t}$ . Otherwise it is called \emph{non-simple material}. } }
\end{dfn}
\begin{Rem}
\rm{{\ From the previous definition, it follows that, in presence of a
simple material, (\ref{pim})--(\ref{pie}) become 
\begin{eqnarray*}
\displaystyle{ {\widetilde{\mathcal{P}}}^i_m(\sigma, P, \widetilde{\gamma}_t)}&=& \displaystyle{{\mathbf{T}%
_2(\sigma, P)\cdot \nabla\widetilde{\mathbf{v}}_t,}}
\\ \displaystyle{ \widetilde{\mathcal{A}}^i_{en}(\sigma, P, \widetilde{\gamma}_t)}&=& \displaystyle{\rho h(\sigma, P) \frac1{\widetilde{\theta}_t}- \mathbf{q}_1(\sigma, P)\cdot \nabla \left[\frac1{\widetilde{\theta}_t}\right],
}
\end{eqnarray*}
where $\mathbf{T}_2$ is the classical Cauchy stress tensor and $\mathbf{q}_1$
is the Fourier heat flux vector. }

\noindent {It should be noticed that the arbitrariness of the virtual process
yields that the triplet $(\mathbf{T}_2, h, \mathbf{q}_1)$ is uniquely
determined when the internal virtual power and entropy action are known. }
{This is in agreement with the well known rational thermodynamic
theories (\cite{Col64Ther}, \cite{ColemanNoll1963}, \cite{ColemanOwen1974}, \cite{noll1958}, \cite{Noll1972}, 
\cite{Truesdell1966}, \cite{Trus69Ratio}, \cite{Truesdell1991}).
%Coleman, Noll, Truesdell). 
}

\noindent {By using the classical notations $(\mathbf{T}, h, \mathbf{q})$
instead of $( \mathbf{T}_2, h, \mathbf{q}_1)$ and $(\mathbf{L},\dot \theta, 
\mathbf{g})$ instead of $( \nabla \mathbf{v}, \dot \theta, \nabla \mathbf{%
\theta})$ (where the dot denotes the time derivative), the relations (\ref{energiainterna2}) and (\ref{entropia2})
become 
\begin{eqnarray}  \label{energiainterna2s}
\frac {d}{dt} e(\sigma_t)&=&{h}(\sigma_t,P(t))+ \frac1\rho {\mathbf{T} }%
(\sigma_t, P(t))\cdot \mathbf{L}(t),
\\
  \label{entropia2s}
\frac {d}{dt}\eta(\sigma_t)&\geq& \frac{{h}(\sigma_t,P(t))}{\theta(t)}+
\frac1{\rho\theta^2(t)} {\mathbf{q} }(\sigma_t, P(t))\cdot \mathbf{g}(t).
\end{eqnarray}
Moreover, introducing the free energy potential 
\begin{equation}  \label{free energy}
\psi= e - \theta \eta,
\end{equation}
(\ref{energiainterna2s}) and (\ref{entropia2s}) yield 
\begin{equation*}  %\label{ free energy 2}
\frac {d}{dt}\psi(\sigma_t)\leq \frac1\rho {\mathbf{T} }(\sigma_t,
P(t))\cdot \mathbf{L}(t) -\frac1{\rho\theta(t)} {\mathbf{q} }(\sigma_t,
P(t))\cdot \mathbf{g}(t)-\dot{\theta}(t)\eta(\sigma_t).
\end{equation*}
}}
\end{Rem}
In the context of the theory of simple materials it is possible to obtain
mathematical models well describing the  thermo-mechanical behavior of
several materials. Here we recall the most elementary examples.

\begin{itemize}
\item \textit{Classical Thermoelastic Systems}

\noindent The state of these materials is given by the deformation gradient $%
\mathbf{F}$ and the absolute temperature $\theta$, i.e. $\sigma=(\mathbf{F},
\theta)$, and the constitutive equations are: 
\begin{equation*} % \label{termoelastico}
\mathbf{T}= {\mathbf{T} }(\sigma)\,,\qquad h= {\mathbf{A} }(\sigma)\cdot {%
\mathbf{L} }+ {B}(\sigma)\dot\theta\,,\qquad \mathbf{q}=\mathbf{K}%
(\sigma)\mathbf{g}.
\end{equation*}
%where ${\mathbf{T}}$, ${\mathbf{A} }$, ${B} $ and $\mathbf{K}$ are
%continuous functions of the state.%$\sigma$.
\item \textit{Newtonian Viscous Fluids}

\noindent The state of these materials is given by the mass density $\varrho 
$ and the absolute temperature $\theta $, i.e. $\sigma =(\varrho ,\theta )$,
and the constitutive equations are: 
\begin{equation*}
\mathbf{T}=[-p(\sigma )+\lambda (\sigma )\mathrm{tr}{\mathbf{D}}]\mathbf{I}%
+2\mu (\sigma ){\mathbf{D}}\,,  \qquad h=A_1(\sigma )\dot{\varrho}+B_1(\sigma )\dot{\theta}\,,\qquad \mathbf{q}={K}%
(\sigma )\mathbf{g},
%\label{fluido viscoso}
\end{equation*}%
where ${\mathbf{D}}=\mathrm{sym}\{\mathbf{L}\}$.
%while $p$, $\lambda $, $\mu  $ ${A_1}$, $B_1$ and ${K}$ are 
%continuous functions of the state.%$\sigma$.
\item \textit{Materials with Fading Memory}

\noindent The state of these materials is given by the histories of the
deformation gradient $\mathbf{F}^t$ and of the absolute temperature $%
\theta^t $, i.e. $\sigma=(\mathbf{F}^t, \theta^t)$ , and the constitutive
equations are: 
\begin{equation*}  %\label{fm}
\mathbf{T}= {\mathbf{T}^{\prime }}(\sigma)\,,\quad h= {\mathbf{A} ^{\prime }}%
(\sigma)\cdot {\mathbf{L} }+ {B}^{\prime }(\sigma)\dot\theta\,,\quad 
\mathbf{q}=\mathbf{K}^{\prime }(\sigma)\mathbf{g}.
\end{equation*}
%where ${\mathbf{T}^{\prime }}$, ${\mathbf{A} ^{\prime }}$, ${B}%^{\prime }$ and $\mathbf{K}^{\prime }$ are 
%continuous functions of the state.%
\item \textit{Materials with internal variables }

\noindent The state of these materials is given by the deformation gradient $%
\mathbf{F}$, the absolute temperature $\theta$ and the internal variables $%
\alpha\in \mathbb{R}^n$, i.e. $\sigma=(\mathbf{F}, \theta,\alpha)$, and the
constitutive equations are: 
\begin{equation*}  %\label{vi}
\mathbf{T}= {\mathbf{T} }^*(\sigma),\;h= {\mathbf{A} ^*}(\sigma)\cdot {%
\mathbf{L} }+ {B}^*(\sigma)\dot\theta,\; \mathbf{q}=\mathbf{K}%
^*(\sigma)\mathbf{g}.
\end{equation*}
%where ${\mathbf{T}}^*$, ${\mathbf{A} ^*}$, ${B}^*$ and $\mathbf{K}^*$ are 
%continuous  functions of the state.%$\sigma$.
\end{itemize}

%%%%%%%%%%%%%%%%%
%MATERIALI NON SEMPLICI
%%%%%%%%%%%%%%%%
\setcounter{equation}{0}
\section{Non-Simple Materials}
As already observed in the Introduction, a significant difference
between simple and non-simple materials is pointed out when the mechanic power and entropy 
action are defined.
For simple materials, the internal mechanical power $\mathcal{P}_{m}^{i}$ is
always expressed by (\ref{Ns1}) and, therefore, the First Law assumes the form\ (\ref{simple1})\ or (\ref%
{simple3}), while the internal  entropy action  $\mathcal{A}_{en}^{i}$ has always the form (\ref{simple17}) and  the Second Law is given by the Clausius Duhem inequality (\ref{simple2}).

For more general materials there are two ways: either (\ref{simple1}) and (\ref{simple2}) must be modified introducing suitable \emph{extra fluxes}, as proposed by Dunn and Serrin \cite{DunnSerrin1985} and M\"uller \cite{Muller1967}, or  the expressions of the internal mechanical powers and of the internal entropy actions
$\mathcal{P}_{m}^{i}$ and $\mathcal{A}_{en}^{i}$  
must be properly derived from the power balance laws. 

If we choose the first option,  the First and Second Law of Thermodynamics are respectively expressed by (\ref{simple6}) and (\ref{simple12}).  It should be noted that the internal powers and actions defined in   (\ref{simple7}) and (\ref{simple171}) are not given by a differential form, but by a hybrid representation containing also  external fluxes.
In any case, we prefer the second option also because it allows us to highlight, without ambiguity,
the contribution of the  internal mechanical power and entropy action  in the variation of the thermodynamic potentials.
\subsection{Second grade materials}
%As an example,
In this subsection  we limit our attention to second grade materials. 
\begin{dfn}
\label{II grado}{\rm A  thermo-mechanical material is called a {\it second
grade material} if 
\begin{equation*}
{\mathbf{T}}_{i+1}={\mathbf{0}}\,,\quad {\mathbf{q}}_{i}={\mathbf{0}}%
\,,\qquad i\geq 3
\end{equation*}%
 for any time $t$ and $x\in \mathcal{B}_{t}$. }
\end{dfn}
In the sequel we will
show how through our theory it is possible to obtain a framework in
agreement with the thermodynamic principles.
\subsubsection{First Law of Thermodynamics}
%When we are in presence of isothermal processes,
For second grade materials, taking in account Definition \ref{II grado},  (\ref{pim}) becomes
\begin{equation*} % \label{secondgrade1}
{\widetilde{\mathcal{P}}}^i_m(\sigma, P, \widetilde{\nu}_t)= \mathbf{T}%
_2(\sigma, P)\cdot \nabla\widetilde{\mathbf{v}}_t + \mathbf{T}_3(\sigma, P)
\cdot\nabla^{(2)}\widetilde{\mathbf{v}}_t,
\end{equation*}
This expression 
fully agrees with the wide literature on these materials. Among all the papers, we recall the one by Fried and Gurtin 
\cite{FriedGurtin2006}. Here the authors, in accordance with \cite%
{Germain1973art}, introduce a general virtual power principle 
%for a second grade material 
by asserting that the internal power has the form 
\begin{equation}
W_{int}=\int_{\Omega }\left( \mathbf{T}_{2}\cdot \nabla \mathbf{v}+\mathbf{T}%
_{3}\cdot \nabla^{(2)} \mathbf{v}\right) dx,  \label{secondgrade3}
\end{equation}%
where $\mathbf{T}_{2}$ and $\mathbf{T}_{3}$ are respectively second and third-order
tensors%
\begin{footnote}{The representation (\ref{secondgrade3}) can be applied to a solid or
liquid flow at small length scale and, with minor modifications, also to second grade nanocristaline elastic materials \cite{Friedgurtin2009}. }\end{footnote}. Moreover they observe that
it is reasonable to modify also the classical kinetic energy 
%$K=\frac12  \rho \mathbf{v}^{2} $ 
and introduce a \emph{generalized kinetic energy} 
% in the following manner 
\begin{equation*}
K^*=\frac{1}{2}(\rho \mathbf{v}^{2}+\beta |\nabla \mathbf{v}|^{2})
%\label{secondgrade4}
\end{equation*}%
to take into account its dependence on the gradient of velocity.

This last observation allows us to study, within this theory, also the
Kirchhoff thermoelastic plate.  Following  \cite{FabLazRiv00Asy}, \cite{Lagn90The},  \cite{LagneseLions1998}, 
 %(see also \cite{Ame03Ex}), 
the constitutive equation for the 
stress tensor is
\begin{equation}\label{Ns5} 
\mathbf{T} =-a\nabla \left(\bigtriangleup u\right)
+c\nabla \theta +\mathbf{T}_{K}
\end{equation}
where $\bigtriangleup$ denotes the Laplace operator, $u$\ is the vertical displacement, $\mathbf{T}_{K}=b\nabla \ddot{u}$  is called kinetic stress and  $a$, $b$ and $c$ are suitable material
constants.

Multiplying the equation of motion 
 \begin{equation*}
\rho \ddot{u}=\nabla \cdot \mathbf{T}+\rho f  %\label{Ns6}
\end{equation*}%
by $\dot{u}$, 
we obtain 
\begin{equation}\frac{d}{dt}\left( \frac{1}{2}\rho \dot{u}^{2}\right) =\nabla \cdot \left( \mathbf{T}\dot{u}\right) -\mathbf{T}\cdot \nabla \dot{u}+\rho f\dot{u} \label{Ns7}
\end{equation}
and making use of the constitutive relation (\ref{Ns5})
%for the stress tensor $\mathbf{T}$,  
the balance equation of the mechanical powers assumes the following form
\begin{eqnarray}
\frac{1}{2}\frac{d}{dt}\left[ \rho \dot{u}^{2}+a\left( \bigtriangleup u\right)
^{2}+b\left( \nabla \dot{u}\right) ^{2}\right] -c\theta \bigtriangleup \dot{u}
 \qquad\qquad\qquad\qquad\qquad\qquad\qquad \notag \\
\qquad\qquad\qquad\qquad\quad =\nabla \cdot \left\{ \left[ -a\nabla \left(\bigtriangleup u\right)
+b\nabla \ddot{u}+c\nabla \theta \right] \dot{u}+(a \bigtriangleup  u-\theta)\nabla \dot{u}%
\right\} +\rho f\dot{u}.  \label{Ns8}
\end{eqnarray}%
By virtue of (\ref{PVP})$_1$
the internal and external mechanical powers are therefore given by%
\begin{equation}  \label{Ns9} 
\mathcal{{P}}_{m}^{i}=\frac{1}{2}\frac{d}{dt}\left[ a\left( \bigtriangleup u\right) ^{2}+b\left( \nabla \dot{u}\right) ^{2}\right] -c\theta \bigtriangleup \dot{u}\,,\qquad
\mathcal{{P}}_{m}^{e}=-\nabla \cdot \mathbf{N}^{\prime }+\rho f\dot{u}\,,
\end{equation}%
where 
\begin{equation}
\mathbf{N}^{\prime }=\left[ a\nabla \left( \bigtriangleup  u\right) -b\nabla 
\ddot{u}-c\nabla \theta \right] \dot{u}-(a \bigtriangleup u-\theta)\,\nabla \dot{u}.
\label{Ns9.2}
\end{equation}%
Consequently, the local form (\ref{energiainterna2}) of the
% new formulation of the 
 First Law  becomes
% is expressed by%
\begin{equation}
\rho \frac{d}{dt}{e}=\frac{1}{2}\frac{d}{dt}\left[ a\left(\bigtriangleup u\right)
^{2}+b\left( \nabla \dot{u}\right) ^{2}\right] -c\theta \bigtriangleup \dot{u}+\rho h  \label{Ns9.1}
\end{equation}%
and it allows us  to better understand the difference between the new
formulation (\ref{simple5}) and the classical one (\ref{simple1})  proposed by Coleman \cite{Col64Ther} and Truesdell and Noll \cite{Trusdell_Noll65,Trus69Ratio}.

If, on the contrary, we want to keep  the classical point of view,  
we  must  modify  (\ref{simple1})
by introducing a suitable extra flux  $\mathbf{N}$ and assume (\ref{simple6}), as suggested by Dunn and Serrin \cite{DunnSerrin1985}.
In fact,  taking into account  (\ref{Ns7})--(\ref{Ns9.2}) and (\ref{simple4}), equation 
(\ref{Ns9.1})  assumes the form
\begin{equation*}
\rho\frac{d}{dt}{e}=
-\frac{d}{dt}\left( \frac{1}{2}\rho \dot{u}^{2}\right) +\rho f\dot{u}-\nabla \cdot \mathbf{N}^{\prime
}+\rho h=
\mathbf{T}\cdot \nabla \dot{u}-\nabla \cdot \left( \mathbf{T}%
\dot{u}+\mathbf{N}^{\prime }\right) -\nabla \cdot \mathbf{q}+\rho r,
%\label{Ns12.2}
\end{equation*}%
which coincides with  (\ref{simple6}) by choosing 
\begin{equation*}
\mathbf{N}= \left( \mathbf{T}\dot{u}+\mathbf{N}^{\prime }\right)
=-a \bigtriangleup  u\,\nabla \dot{u}. %\label{Ns13}
\end{equation*}%
Let us now consider a thermoelastic plate with
memory characterized by the following constitutive equation%s%
\begin{equation}  \label{Ns13.1}
\mathbf{T}(t)=-\nabla \left[ \int_{0}^{\infty}\!\!\!\!C^{\prime
}(s) \bigtriangleup  u^t(s)ds+C_{0} \bigtriangleup  u(t)\right]
+c\nabla \theta (t),
\end{equation}
where  the scalar kernel $C^{\prime }$\
is a suitable smooth function and $u^t(s)=u(t-s)$.

Reasoning in the same way as in the previous model, we evaluate the terms in the right-hand side of (\ref{Ns7}) making use of (\ref{Ns13.1}) and  obtain
\begin{eqnarray*}
\frac{d}{dt}\left( \frac{1}{2}\rho \dot{u}^{2}(t)\right) &+&\left[\int_{0}^{\infty}\!\!\!\!C^{\prime }(s) 
\bigtriangleup  u^t(s)ds\, 
-c\theta (t)\right]  \bigtriangleup  \dot{u}(t)+\frac{C_{0}}{2}%
\frac{d}{dt}\left[  \bigtriangleup  u(t)\right] ^{2} \qquad 
 \\
&&=\nabla \cdot \left( \left\{ -\nabla \left[ \int_{0}^{\infty}\!\!\!\!C^{\prime }(s) \bigtriangleup  u^t(s)ds+C_{0} \bigtriangleup  u(t)\right] +c\nabla
\theta(t) \right\} \dot{u}(t)\right.   \notag \\
&&\quad \left. +\left[ \int_{0}^{\infty}\!\!\!\!C^{\prime }(s)\bigtriangleup
u^t(s)ds+C_{0} \bigtriangleup  u(t)-c\theta (t)\right] \nabla \dot{u}(t)\right) +\rho
f(t)\dot{u}(t). %\label{Ns13.3}
\end{eqnarray*}%
Hence, it follows that the internal and external mechanical powers are given by%
$$
\mathcal{{P}}_{m}^{i}(t)=\left[\int_{0}^{\infty}\!\!\!\!C^{\prime }(s) 
\bigtriangleup  u^t(s)ds\, 
-c\theta (t)\right]  \bigtriangleup  \dot{u}(t)+\frac{C_{0}}{2}%
\frac{d}{dt}\left[  \bigtriangleup  u(t)\right] ^{2} \,,\qquad
\mathcal{{P}}_{m}^{e}(t)=-\nabla \cdot \mathbf{N}^{\prime }(t)+\rho f(t)\dot{u}(t)%
$$
where 
\begin{eqnarray*}% \label{Ns13.6}
\mathbf{N}^{\prime }(t)&=&\left\{ \nabla \left[ \int_{0}^{\infty}\!\!\!\!C^{\prime
}(s) \bigtriangleup  u^t(s)ds+C_{0} \bigtriangleup  u(t)\right] -c\nabla \theta(t) \right\} 
\dot{u} (t) \\&&-\left[ \int_{0}^{\infty}\!\!\!\!C^{\prime }(s) \bigtriangleup  u^t(s)ds+C_{0}\bigtriangleup
u(t)-c\theta (t)\right] \nabla \dot{u}(t).
\end{eqnarray*}
Now, using the formulation  (\ref{simple5}) of the First Law, we could define the internal heat power as follows%
\begin{equation*}  % \label{Ns13.5}
\rho h (t)
=\rho \frac{d}{dt}{e}(t)%
-\left[\int_{0}^{\infty}\!\!\!\!C^{\prime }(s) 
\bigtriangleup  u^t(s)ds\, 
-c\theta (t)\right]  \bigtriangleup  \dot{u}(t)-\frac{C_{0}}{2}%
\frac{d}{dt}\left[  \bigtriangleup  u(t)\right] ^{2} .
\end{equation*}
Finally,  by choosing
\begin{equation*}
\mathbf{N}(t)=
%\mathbf{T}\dot{u}+\mathbf{N}^{\prime }=
-\left[ \int_{0}^{\infty
}C^{\prime }(s) \bigtriangleup  u^t(s)ds+C_{0} \bigtriangleup  u(t)-c\theta (t)\right]
\nabla \dot{u}(t),  %\label{Ns13.7}
\end{equation*}
we can use also  (\ref{simple6}) as local form of the First Law.

Other interesting examples, for which the classical representation of the
power, used in the First Law, does not coincide with the one following from
a correct balance between the internal and external powers, meet in
electromagnetism, in the study of phase transitions and also in other
physical systems, as those describing processes of phase separation, for
example, by means of Cahn-Hilliard's equation \cite{CahHil58Fre}.

Firstly, we consider %an 
a rigid electromagnetic system for which the classical
expression for the First Law is given by%\footnote{conduttore rigido e $\rho=1$}%
\begin{equation}
\frac{d}{dt}{e}=\mathbf{\dot{B}}\cdot \mathbf{H}+\mathbf{\dot{D}}\cdot \mathbf{E}+%
\mathbf{J}\cdot \mathbf{E}-\nabla \cdot \mathbf{q}+r,  \label{Ns14}
\end{equation}%
where $\mathbf{E}$, $\mathbf{H}$, $\mathbf{D}$,\ $\mathbf{B}$,\ $\mathbf{J}$%
\ are\ the electric and magnetic fields,\ the electric displacement, the
magnetic induction and the current density, respectively. 

Equation  (\ref{Ns14}) together with  (\ref{simple4}) leads to the classical definition of internal electromagnetic power
\begin{equation}\label{Pie}
\mathcal{{P}}_{el}^{i}=\mathbf{\dot{B}}\cdot \mathbf{H}+\mathbf{\dot{D}}\cdot \mathbf{E}+
\mathbf{J}\cdot \mathbf{E}
\end{equation}
and thermal power 
\begin{equation}\label{acca}
h=\frac{d}{dt}{e}-\mathbf{\dot{B}}\cdot \mathbf{H}-\mathbf{\dot{D}}\cdot \mathbf{E}-
\mathbf{J}\cdot \mathbf{E}.
\end{equation}
These relations hold
for a large class of electromagnetic materials, such as classical dielectrics, conductors
characterized by the Ohm Law, materials with memory, etc. \cite{Fab_Laz_Mor_2003}.
In analogy with the mechanical systems, these materials are still called
simple. However there exist meaningful examples of fairly common materials, for which (%
\ref{Ns14}) must be modified by  changing the internal electromagnetic power $\mathcal{{P}}_{el}^{i}$ or by introducing an extra flux
as follows%
\begin{equation}
\frac{d}{dt}{e}=\mathbf{\dot{B}}\cdot \mathbf{H}+\mathbf{\dot{D}}\cdot \mathbf{E}+%
\mathbf{J}\cdot \mathbf{E}-\nabla \cdot \mathbf{N}-\nabla \cdot \mathbf{q}+r.
\label{Ns21}
\end{equation}%
To this purpose we recall that, as already shown in \cite{bib:fabmem}, in presence of superconducting materials it is necessary to use (\ref{Ns21}) with an
extra flux proportional to the superconducting current.

Here we consider a dielectric with quadrupoles described by the
following constitutive equations%
\begin{eqnarray}
\mathbf{D} &=&\varepsilon _{0}\mathbf{E}-\varepsilon _{1} \bigtriangleup  \mathbf{E}-\varepsilon
_{2}\nabla \left[ \nabla \cdot \mathbf{E}\right] ,  \notag \\
\mathbf{B}&=&\mu \mathbf{H},  \label{Ns15} \\
\mathbf{J}&=&\mathbf{0},  \notag
\end{eqnarray}%
where $\varepsilon _{0}$, $\varepsilon _{1}$,\ $\varepsilon _{2}$\ are
suitable  constants characteristic of the dielectric.

The evolution of the electromagnetic field is governed by the well known Maxwell equations which, in absence of currents,  are%
\begin{eqnarray*}
\mathbf{\dot{D}} &=&\nabla \times \mathbf{H}, \\%\label{Ns16} \\
\mathbf{\dot{B}} &=&-\nabla \times \mathbf{E}.  %\label{Ns16.1}
\end{eqnarray*}%
An immediate consequence of the previous equations is the balance law for the electromagnetic  power, known as Poynting theorem,
\begin{equation*}
\mathbf{\dot{D}}\cdot \mathbf{E}+\mathbf{\dot{B}}\cdot \mathbf{H}=\nabla
\times \mathbf{H}\cdot \mathbf{E}-\nabla \times \mathbf{E}\cdot \mathbf{H}=\mathbf{-}\nabla \cdot \left( \mathbf{E}\times \mathbf{H}\right) % \label{Ns17.0.1}
\end{equation*}%
which,
taking into account the constitutive equations (\ref{Ns15}), becomes %
\begin{equation*}
\frac12\frac{d}{dt}\left[ \mu |{\mathbf{H}|^{2}}+\varepsilon _{0}{
|\mathbf{E}|^{2}}+\varepsilon _{1}|{\nabla \mathbf{E}| ^{2}
}+\varepsilon _{2}{|\nabla \cdot \mathbf{E}|^{2}}\right]  =-\nabla \cdot \left[ \mathbf{E}\times \mathbf{H}%
-\varepsilon _{1}\nabla \mathbf{\dot{E}\,E}-\varepsilon _{2}\nabla \cdot
\mathbf{\dot{E}\,E}\right] .  %\label{Ns17.1}
\end{equation*}%
 Hence the  expressions of the internal and external electromagnetic powers are
\begin{eqnarray*}
\mathcal{{P}}_{el}^{i} &=&\frac12\frac{d}{dt}\left[ \mu |{\mathbf{H}|^{2}}+\varepsilon _{0}{
|\mathbf{E}|^{2}}+\varepsilon _{1}|{\nabla \mathbf{E}| ^{2}
}+\varepsilon _{2}{|\nabla \cdot \mathbf{E}|^{2}}\right],\\
%\label{Ns18} \\
\mathcal{{P}}_{el}^{e} &=&-\nabla \cdot \left[ \mathbf{E}\times
\mathbf{H}-\varepsilon _{1}\nabla \mathbf{\dot{E}\,E}-\varepsilon _{2}\nabla
\cdot \mathbf{\dot{E}\,E}\right] .  \label{Ns19}
\end{eqnarray*}
Consequently, the local form (\ref{energiainterna2}) of the
 First Law,  with $\mathcal{{P}}_{el}^{i} $ instead of  $\mathcal{{P}}_{m}^{i} $,   becomes
\begin{equation*}
\frac{d}{dt}{e}=\frac{1}{2}\frac{d}{dt}\left[ \mu \mathbf{H}^{2}+\varepsilon _{0}%
\mathbf{E}^{2}+\varepsilon _{1}\left( \nabla \mathbf{E}\right)
^{2}+\varepsilon _{2}(\nabla \cdot \mathbf{E})^{2}\right] +h
%\label{Ns20}
\end{equation*}%
and it allows us to give a properly definition of
the thermal power $h$
which does not agree with the classical definition (\ref{acca}).

Finally, it is easy to show that the relation (\ref{Ns21}) is satisfied by choosing
\begin{equation*}
\mathbf{N}=-\varepsilon _{1}\nabla \mathbf{\dot{E}\,E}-\varepsilon
_{2}\nabla \cdot \mathbf{\dot{E}\,E.}  %\label{Ns21.1}
\end{equation*}
Therefore, this example confirms that the local  form (\ref{energiainterna2}) of the
 First Law (with $\mathcal{{P}}_{el}^{i} $ instead of  $\mathcal{{P}}_{m}^{i} $)   has a universal character; on the contrary, (\ref{acca}) must be adapted in function of the material under  consideration.

Let us now consider an example  related to the phase transition in a
binary mixture characterized by
the Cahn-Hilliard equation
\begin{footnote}
{Among the wide literature on 
%the study of 
the Cahn-Hilliard equation we recall, for example, \cite{Morris_1973} and \cite{Gurtin_1989} that set precedents.}
\end{footnote}%
\begin{equation}
\dot{c}=\nabla \cdot [M(c)\nabla \mu(c)],  \label{CH1}
\end{equation}
where $c \in [-1,1]$  denotes the relative concentration of one of the two components, the   \emph{mobility} $M$  represents the density with which
the two phases are mixed and
 $\mu $ is the\emph{ chemical potential}. 

The   {mobility}  $M$ is  a nonnegative function 
and the {chemical potential} $\mu$ is given by%
\begin{equation}
\mu(c) =-\gamma \bigtriangleup c+\theta _{0}F^{\prime }(c)+\theta G^{\prime }(c),  \label{CH2}
\end{equation}
where $\gamma$ is a positive constant, $\theta _{0}$ is the temperature of the transition, while $G$ and $F$ are the usual polynomials assumed for a second order transition, i. e.
\begin{equation}
\label{CH3}
G(c) = \beta \frac{c^2}{2}, \qquad F(c) = \beta \left( \frac{c^4}{4} - \frac{c^2}{2} \right), \quad \beta > 0.
\end{equation}
Therefore, from (\ref{CH1})--(\ref{CH3}), we obtain
\begin{equation}
\label{CH4}
\dot{c}=\nabla \cdot \left[ M(c)\nabla (-\gamma \bigtriangleup c+\theta _{0}F^{\prime
}(c)+\theta G^{\prime }(c))\right].
\end{equation}
By
multiplying (\ref{CH4}) by the potential $\mu$, we have the equation
for the power%
\begin{equation*}
\dot{c}\mu (c)=\nabla \cdot \left[ M(c)\mu (c)\nabla \mu (c)\right] -M(c)%
\left[ \nabla \mu (c)\right] ^{2}\mathbf{,}  %\label{Ns29}
\end{equation*}%
which, by virtue of (\ref{CH3}), can be rewritten as%
\begin{equation*}
\theta _{0}\dot{F}(c)+\theta \dot{G}(c)+\frac{\gamma }{2}\frac{d}{dt}%
\left( \nabla c\right) ^{2}+M(c)\left[ \nabla \mu (c)\right] ^{2}  =\nabla \cdot \left[ \gamma \dot{c}\nabla c+M(c)\mu (c)\nabla \mu
(c)\right]. % \label{Ns29.1}
\end{equation*}%
This  relation allows us to define  the internal and external chemical powers of the mixture
\begin{equation} \label{CH6}
\begin{array}{ll}
   {\cal P}^i_c    & =  \displaystyle{ \theta _{0}\dot{F}(c)+\theta \dot{G}(c)+\frac{\gamma }{2}\frac{d}{dt}%
\left( \nabla c\right) ^{2}+M(c)\left[ \nabla \mu (c)\right] ^{2} \,, } \\
    {\cal P}^e_c  & =   \displaystyle{ \nabla \cdot \left[ \gamma \dot{c}\nabla c+M(c)\mu (c)\nabla \mu
(c)\right].}
\end{array}
\end{equation}
If we put (\ref{CH6}) into the heat equation
\begin{equation*}
\frac{d}{dt}{e}-\mathcal{{P}}_{c}^{i}=-\nabla \cdot \mathbf{q}+r, % \label{Ns28}
\end{equation*}%
we get
\begin{equation}
\frac{d}{dt}{e}=\dot{c}\mu (c)+\gamma \nabla \cdot (\dot{c}\nabla c)+M(c)\left[
\nabla \mu (c)\right] ^{2}-\nabla \cdot \mathbf{q}+r.  \label{Ns33}
\end{equation}%
It follows that if we want to use 
the usual expression $\dot{c}\mu (c)$ for the chemical power,  even in this case, it is necessary to introduce 
 the extra flux%
\begin{equation*}
\mathbf{N}=-\gamma \dot{c}\nabla c. % \label{Ns33.1}
\end{equation*}
We conclude this subsection by observing that 
if we take for the internal energy the  constitutive equation
\begin{equation*}
e(\theta, c,\nabla  c) =\theta _{0}F(c)+\frac{\gamma }{2}\left( \nabla c\right) ^{2}+\tilde{e}%
(\theta ),
\end{equation*}%
then equation (\ref{Ns33}) assumes the following more usual form %
\begin{equation*}
\frac{d}{d\theta}\tilde{e}(\theta)\dot{\theta} =\theta \dot{G}(c)+M(c)\left[ \nabla \mu (c)%
\right] ^{2}-\nabla \cdot \mathbf{q}+r. % \label{Ns31}
\end{equation*}

A similar argument can be also applied to the binary mixture of two incompressible fluids  studied  by 
Lowengrub and Truskinovsky  in \cite{Truski}.

\subsubsection{Second Law of Thermodynamics}

In this subsection we consider only non-simple materials in the heat flux, in other words, materials for which the constitutive equation for the heat flux depends also on the gradients of order higher than the first in the temperature. 

For second grade materials, taking in account Definition \ref{II grado},  the internal virtual entropy action (\ref{pie}) becomes
\begin{equation}  \label{secondgrade2}
\widetilde{\mathcal{A}}^i_{en}(\sigma, P, \widetilde{\gamma}_t)= \rho
h(\sigma, P) \frac1{\widetilde{\theta}_t}- \mathbf{q}_1(\sigma, P)\cdot
\nabla\left[\frac1{\widetilde{\theta}_t}\right] - \mathbf{q}_2(\sigma,
P)\cdot \nabla^{(2)}\left[\frac1{\widetilde{\theta}_t}\right].
\end{equation}%
Inequalities  (\ref{simple2}) and (\ref{simple12}) or, equivalently, 
\begin{eqnarray}
\rho \frac{d}{dt}{\eta}\geq& 
\displaystyle{\rho \frac{h}{\theta }+\frac{1}{%
\theta ^{2}}\mathbf{q}\cdot \nabla \theta} \quad\qquad\qquad&\hbox{for simple materials} ,\label{Ns38}\\
\rho\frac{d}{dt}{\eta}\geq& \displaystyle{\rho \frac{h}{\theta }+\frac{1}{\theta ^{2}}\mathbf{q}%
\cdot \nabla \theta -\nabla \cdot \mathbf{\Phi }^{\prime }} \quad&\hbox{for non-simple materials} , \label{Ns39}
\end{eqnarray}
show that the Second Law assumes different expressions for
simple or non-simple materials. 
	
Reasoning in  same way as for  
the First Law, we suggest to claim the Second Principle in a unique general
form trough the inequality (\ref{simple191}).   

For simple materials the internal entropy action $%
\mathcal{A}_{en}^{i}$ is always given by the differential form of the right-hand side of (\ref{Ns38}), while the right-hand side of (\ref{Ns39}) 
cannot be taken as expression of  $\mathcal{A}_{en}^{i}$ for non-simple materials, because of the presence of the external  flux $\nabla \cdot  \mathbf{\Phi }^{\prime }$.

We now present an example  \cite {AmFaGo} that helps to better understand the motivations of our suggestion.
Let us  consider a rigid heat conductor, whose constitutive equation is
given by
\begin{equation}
\mathbf{q}(t)=\mathbf{q}_{1}(\theta(t),{ \bar{g}}^{t}) -\nabla\!\cdot \mathbf{q}_{2}(\theta(t), {\bar{g}}^{t}, \nabla{\bar{g}}^{t}) \label{f.2g}
\end{equation}%
where   $\mathbf{q}_{1}$ is a vector,   $\mathbf{q}_{2}$  is a second order tensor and $\mathbf{%
\bar{g}}^{t}$ is\ the integrated history of $\nabla \theta$, that is%
\begin{equation*}
\mathbf{\bar{g}}^{t}(s)=\int_{t-s}^{t}\mathbf{g}(\lambda )d\lambda\,. % \label{f.1}
\end{equation*}
The functional (\ref{f.2g}) is a typical constitutive equation of a second
grade material; therefore, taking into account  %the representation 
(\ref{secondgrade2}), the internal entropy action becomes
\begin{equation*}
\mathcal{A}_{en}^{i}= \frac{1}{\theta^2 } \left[\theta h+\mathbf{q}_{1}\cdot \nabla \theta
+\mathbf{q}_{2}\cdot \nabla^{(2)} \theta-\frac2\theta
\mathbf{q}_{2}\nabla\theta\cdot \nabla\theta \right]. % \label{f.3}
\end{equation*}%
Consequently, the First and Second Thermodynamic Laws  have the following expressions%
\footnote{We can assume $\rho=1$ without loss of generality.}%
\begin{eqnarray}
\frac{d}{dt}{e}&=&h\, ,  \label{f.4}
\\
\frac{d}{dt}{\eta}&\geq& \ \frac{1}{\theta^2 } \left[\theta h+\mathbf{q}_{1}\cdot \nabla \theta
+\mathbf{q}_{2}\cdot \nabla^{(2)} \theta-\frac2\theta
\mathbf{q}_{2}\nabla\theta\cdot \nabla\theta \right]\, . \label{f.5}
\end{eqnarray}
We observe that, choosing $$ \mathbf{\Phi }^{\prime }=-\frac1 {\theta^2}\ {\mathbf{q}_{2}}\nabla {\theta},$$ the inequality  (\ref{Ns39})  coincides with  (\ref{f.5}).

Finally, we assume the following constitutive equations for the internal energy and  heat flux
\begin{equation*}
\begin{array}{lll}
\displaystyle{     e(\theta(t), \mathbf{\bar{g}}^{t}, \nabla\mathbf{\bar{g}}^{t})}&=&\displaystyle{   \tilde{e}(\theta(t) )+b(\mathbf{\bar{g}}^{t}, \nabla\mathbf{\bar{g}}^{t})},    \\
   \displaystyle{      \mathbf{q}_{1}(\theta(t), \mathbf{\bar{g}}^{t})}&=&\displaystyle{-   {\theta(t)} \int_{0}^{+\infty }K_{1}^{\prime }(s)%
\mathbf{\bar{g}}^{t}(s)ds}\,,
\\
\displaystyle{   \mathbf{q}_{2}(\theta(t), \mathbf{\bar{g}}^{t}, \nabla\mathbf{\bar{g}}^{t})}&=&\displaystyle{  - {\theta(t)}
\int_{0}^{+\infty }K_{2}^{\prime }(s)\nabla \mathbf{\bar{g}}^{t}(s)ds},
  \end{array}
\end{equation*}%
with $\tilde{e}$ positive and homogeneous function of $\theta $, $K_{1}$\ and\ $K_{2}$\  smooth positive functions.

Then the internal entropy action becomes
\begin{eqnarray*}
\mathcal{A}_{en}^{i}(t)&=& \frac{1}{\theta(t) } \left[ h(t)-\int_{0}^{+\infty }K_{1}^{\prime }(s)%
\mathbf{\bar{g}}^{t}(s)ds\cdot \nabla \theta(t)
-\int_{0}^{+\infty }K_{2}^{\prime }(s)\nabla \mathbf{\bar{g}}^{t}(s)ds\cdot \nabla^{(2)} \theta(t)\right] \\ &&
+
\frac{2}{\theta^2(t)} \int_{0}^{+\infty }K_{2}^{\prime }(s)\nabla \mathbf{\bar{g}}^{t}(s)ds\nabla \theta(t)\cdot \nabla \theta(t) % \label{f.3}
\end{eqnarray*}%
and the free energy potential (\ref{free energy})  can be expressed as a sum of two terms
$$\psi(\theta(t), \mathbf{\bar{g}}^{t}, \nabla\mathbf{\bar{g}}^{t})= \psi _{1}(\theta (t))+\psi _{2}(\mathbf{\bar{g}}^{t}, \nabla\mathbf{\bar{g}}^{t}),
$$
where 
\begin{equation*}
\psi _{1}(\theta )=\tilde{e}(\theta )-\theta \int _0^\theta\frac1\tau \frac{d}{d\tau}\tilde{e}(\tau )d\tau 
\end{equation*}%
and the functional $\psi _{2}$ satisfies the
inequality%
\begin{eqnarray*}
\frac{d}{dt}\psi _{2}(\mathbf{\bar{g}}^{t}, \nabla\mathbf{\bar{g}}^{t}))&\leq& \int_{0}^{+\infty }K_{1}^{\prime }(s)%
\mathbf{\bar{g}}^{t}(s)ds\cdot \nabla \theta (t)+ \int_{0}^{+\infty
}K_{2}^{\prime }(s)\nabla \mathbf{\bar{g}}^{t}(s)ds\cdot \nabla^{(2)} \theta (t) \\
 &&
-
\frac{2}{\theta(t)} \int_{0}^{+\infty }K_{2}^{\prime }(s)\nabla \mathbf{\bar{g}}^{t}(s)ds\nabla \theta(t)\cdot \nabla \theta(t) .% \label{f.3}
\end{eqnarray*}%
\subsection{Other non-simple materials}
It is worth to observe that there exist material systems, whose constitutive equations present higher order
gradients, for which the internal mechanical virtual  power and internal virtual entropy action are  not covered by the expressions (\ref{pim}) and  (\ref{pie}). 

For this purpose, 
let us  consider the model proposed by Guyer and Krumhansl \cite{Guyer-Krumhansl1966b} 
for the heat flux and recently studied in   \cite{Cimmelli2007,triani-papenfuss-cimmelli2008}.
Since the constitutive law for heat flux is given by %
\begin{equation}
\mathbf{\dot{q}}+\frac{1}{\tau _{R}}\mathbf{q}=-c_{V}\nabla \theta +\tau _{N}%
\left[  \bigtriangleup  \mathbf{q}+2\nabla \left( \nabla \cdot \mathbf{q}\right) %
\right] ,%\quad \tau _{R}>0,\;\tau _{N}>0, 
\label{Ns40}
\end{equation}%
it is easy to show that for  this material  the internal virtual  entropy  action cannot be
represented in the form (\ref{pie}).
Anyway,  we can still define the internal entropy action. In fact, if we choose
\begin{equation*}
c_{V}=\frac{c_{0}}{\theta ^{2}},\quad c_{0}>0, % \label{Ns41}
\end{equation*}%
and multiply  (\ref{Ns40}) by $\mathbf{q}$,  we obtain 
\begin{equation}\label{f.6}
\frac{1}{\theta ^{2}}\mathbf{q}\cdot \nabla \theta=
-\frac{1}{c_{0}}\left[\frac12\frac{d}{dt}\mathbf{q}%
^{2}+\frac1{\tau _{R}}\mathbf{q}^{2}+{\tau _{N}}
\left( \nabla \mathbf{q}\right) ^{2}+2\left( \nabla \cdot \mathbf{q}\right)
^{2}\right] +\frac{\tau _{N}}{c_{0}}\nabla \cdot \left( \nabla \mathbf{q}\,\mathbf{q}%
+2\nabla \cdot \mathbf{q\,q}\right).
\end{equation}
The heat equation for a rigid conductor together with (\ref{f.6}) yields
\begin{eqnarray*}
\frac1{\theta}\frac{d}{dt}{e}-\frac{1}{2c_{0}}\frac{d}{dt}\mathbf{q}%
^{2}-\frac{1}{c_{0}\tau _{R}}\mathbf{q}^{2}&-&\frac{\tau _{N}}{c_{0}}\left[
\left( \nabla \mathbf{q}\right) ^{2}+2\left( \nabla \cdot \mathbf{q}\right)
^{2}\right]  \\
&& =-\nabla \cdot \left(\frac{\mathbf{q}}{\theta }\right)+\rho \frac{r}{\theta }-%
\frac{\tau _{N}}{c_{0}}\nabla \cdot \left( \nabla \mathbf{q}\,\mathbf{q}%
+2\nabla \cdot \mathbf{q\,q}\right) . \label{Ns41.1}
\end{eqnarray*}%
Hence, the expressions of the internal and external entropy actions are given by
\begin{eqnarray*}
\mathcal{{A}}_{en}^{i}&=&\frac1{\theta}\frac{d}{dt}{e}-\frac{1}{2c_{0}%
}\frac{d}{dt}\mathbf{q}^{2}-\frac{1}{c_{0}\tau _{R}}\mathbf{q}^{2}-\frac{%
\tau _{N}}{c_{0}}\left[ \left( \nabla \mathbf{q}\right) ^{2}+2\left( \nabla
\cdot \mathbf{q}\right) ^{2}\right] , % \label{Ns41.2}
\\
\mathcal{{A}}_{en}^{e}&=&-\nabla \cdot \left(\frac{\mathbf{q}}{\theta }\right)+\frac{r%
}{\theta }-\nabla \cdot  \mathbf{\Phi }_{0}^{\prime }, % \label{Ns41.3}
\end{eqnarray*}%
where 
\begin{equation*}
 \mathbf{\Phi }_{0}^{\prime }=\frac{\tau _{N}}{c_{0}}\left( \nabla \mathbf{q}%
\,\mathbf{q}+2\nabla \cdot \mathbf{q\,q}\right)  % \label{Ns42}
\end{equation*}
and the Second Law of Thermodynamic (\ref{simple191}) becomes
\begin{equation}
\frac{d}{dt}{\eta}\geq \frac1{\theta}\frac{d}{dt}{e}-\frac{1}{2c_{0}}\frac{d}{dt}%
\mathbf{q}^{2}-\frac{1}{c_{0}\tau _{R}}\mathbf{q}^{2}-\frac{\tau _{N}}{c_{0}}%
\left[ \left( \nabla \mathbf{q}\right) ^{2}+2\left( \nabla \cdot \mathbf{q}%
\right) ^{2}\right] . \label{Ns43}
\end{equation}%
The previous inequality imposes restrictions on the
constitutive equation (\ref{Ns40}). In fact, if we choose
 the following expression for the entropy%
\begin{equation*}
\eta = \int _0^\theta\frac1\tau \frac{\partial {e}}{\partial\tau}d\tau  -\frac{1}{2c_{0}}\mathbf{q}^{2}\,,
\end{equation*}
then from (\ref{Ns43}) we obtain that the inequality%
\begin{equation*}
\frac{1}{c_{0}\tau _{R}}\mathbf{q}^{2}+\frac{\tau _{N}}{c_{0}}\left[ \left(
\nabla \mathbf{q}\right) ^{2}+2\left( \nabla \cdot \mathbf{q}\right) ^{2}%
\right] \geq 0
\end{equation*}%
 must be satisfied for any field $\mathbf{q}$ and therefore the constitutive coefficients $\tau _{R}$ and $\tau _{N}$ must be positive.

\begin{acknowledgements} 
This research was performed under the auspices of GNFM (INDAM).
\end{acknowledgements}

\end{document}